\documentclass[nofootinbib,superscriptaddress,twocolumn,showpacs,preprintnumbers,amsmath,amssymb]{revtex4-2}

\usepackage{orcidlink}
\usepackage{listings}
\usepackage{graphicx}
\usepackage{grffile}
\usepackage{slashed}
\usepackage{bbm}
\usepackage{footmisc}
\usepackage{multirow}

\usepackage{hyperref}
\usepackage{amssymb}
\usepackage{amsmath}
\usepackage{xcolor}
\usepackage{xspace}

\usepackage{cleveref}

\usepackage{datetime}
\usepackage{color,fancybox}

\newcommand{\bea}{\begin{eqnarray}}
\newcommand{\eea}{\end{eqnarray}}

\newcommand{\fig}[1]{Fig.~\ref{#1}}

\def\3l{\ensuremath{\ell \nu \ell \nu \ell \nu}\xspace}

\newcommand{\Vbfnlo}{\textsc{Vbfnlo}}
\newcommand{\Herwig}{\mbox{\textsf{Herwig}}~}

\begin{document}

\title{NLO QCD parton shower matching for $p p \rightarrow e^{+} \nu_e \mu^{-} \bar{\nu}_{\mu} \gamma + X$}

\author{Ivan~Rosario\, \orcidlink{0000-0002-9445-530X}}
\email{ivan.rosario@ific.uv.es}
\affiliation{Theory Division, IFIC, University of Valencia-CSIC, E-46980
  Paterna, Valencia, Spain}
  
\author{Francisco~Campanario}
\email{francisco.campanario@ific.uv.es}
\affiliation{Theory Division, IFIC, University of Valencia-CSIC, E-46980
  Paterna, Valencia, Spain}

\author{Simon~Pl{\"a}tzer}
\email{simon.plaetzer@univie.ac.at}
\affiliation{Institute of Physics, NAWI Graz, University of Graz, Universitätsplatz 5, A-8010 Graz, Austria }
\affiliation{Particle Physics, Faculty of Physics, University of Vienna, Boltzmanngasse 5, A-1090 Wien, Austria }

\begin{abstract}
We present the implementation of a new interface in \Vbfnlo{} 3.0 supporting all di-boson and
tri-boson processes with fully leptonic final states, enabling NLO+PS matched calculations.
To demonstrate its capabilities, we study parton shower effects in the tri-boson production
process $p p \rightarrow e^{+} \nu_e \mu^{-} \bar{\nu}_{\mu} \gamma + X$  using Herwig 7.3
with NLO QCD amplitudes from \Vbfnlo{} 3.0. We estimate uncertainties from scale variations and
analyze the impact of generation-level cuts on parton shower events. This study showcases the
new interface's potential and provides insights into the interplay between fixed-order
calculations and parton shower effects in multi-boson production processes, crucial for
precision measurements and BSM searches at the LHC and future colliders.
\end{abstract}

\maketitle

\section{Introduction}
\label{sec:intro}

The Standard Model (SM) stands as our most comprehensive and precise framework
for understanding particle physics. At its core, it is a Quantum Field Theory governed
by the principle of local gauge invariance, which dictates the nature of particle interactions.
The discovery of the Higgs boson in 2012 by the CMS \cite{CMS:2012qbp} 
and ATLAS \cite{ATLAS:2012yve} collaborations at the Large Hadron Collider (LHC) completed the experimental 
verification of SM particle spectrum, resolving the
long-standing challenge of reconciling gauge boson masses
with local gauge invariance throught the mechanism of
spontanous symmetry breaking.

After the discovery of the Higgs boson, 
the focus of particle physics has shifted to precision measurements of SM 
parameters and searches for physics beyond the SM (BSM). 
High-mass or weakly interacting particles may induce subtle effects on 
distributions and cross-sections at LHC energies, necessitating high precision
in both experimental and theoretical approaches.
Multi-boson production processes offer a promising avenue for probing these deviations.
As the SM gauge principle fully determines electroweak boson couplings,
these processes facilitate systematic studies of BSM effects through frameworks
like Effective Field Theories (EFTs).

Accurate comparison between theoretical predictions and 
experimental data relies heavily on event generators. These 
tools combine perturbative matrix element 
calculations with parton showers and non-perturbative effects to 
compute observable quantities. In this context, understanding the
effects of parton showers on next-to-leading order (NLO) quantum 
chromodynamics (QCD) calculations becomes crucial for precise 
predictions and uncertainty estimation.

\begin{figure}
  \includegraphics[width=0.46\textwidth]{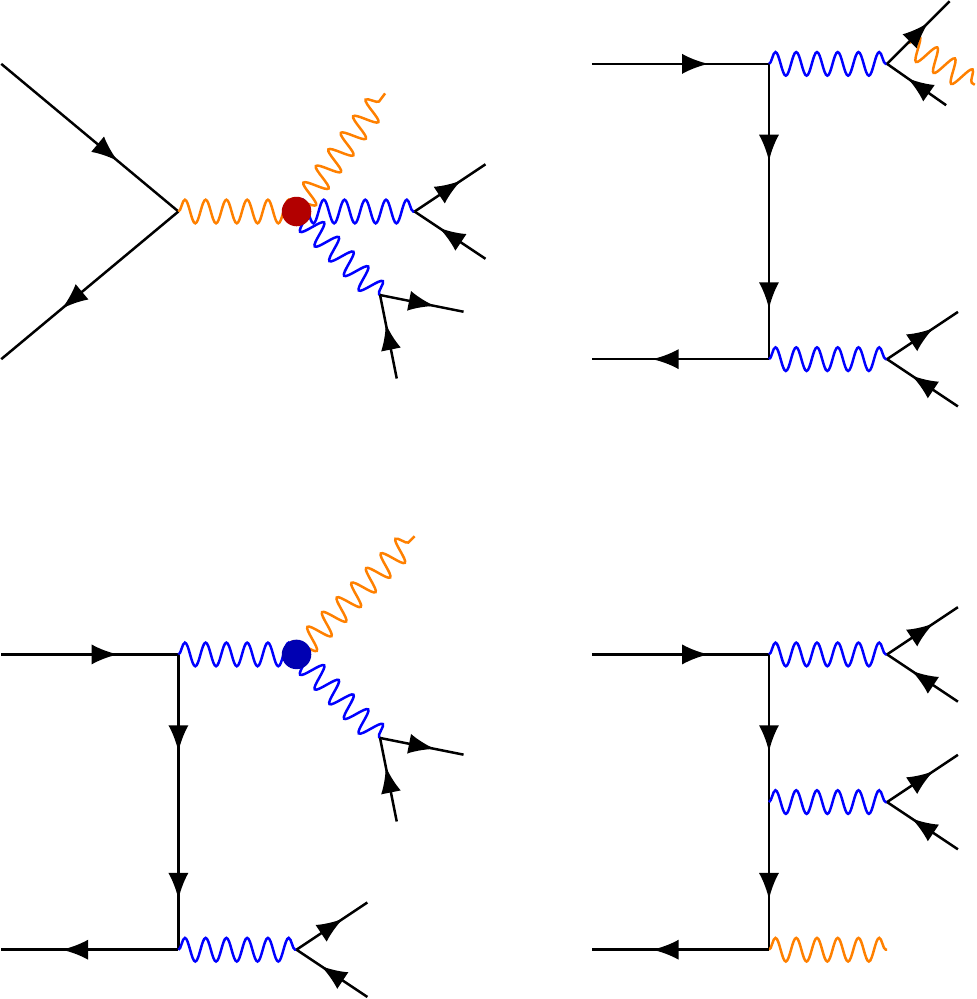}
  \caption{Representative set of LO diagrams for tri-boson production processes. In the top left diagram,
    we can see a quartic gauge coupling (in red) where a s-channel photon decays into
    the two $W$ and another photon.
    In the bottom left is a diagram with a triple gauge coupling (in blue), while the two right diagrams
    do not include any boson-boson coupling.}
  \label{fig:tri-boson-diagrams}
\end{figure}

In this work, we study the parton shower effects on NLO QCD matrix elements in the
tri-boson production process $p p \rightarrow e^{+} \nu_e \mu^{-} \bar{\nu}_{\mu} \gamma$
($W^{+}W^{-}\gamma$) using the \Herwig 7.3 event generator~\cite{Bahr:2008pv, Bellm:2015jjp} and \Vbfnlo{} 3.0~\cite{Arnold:2008rz, Baglio:2011juf, Baglio:2014uba, baglio2024release} as the NLO QCD
amplitude provider. The communication between both programs is done through the BLHA
interface.
Tri-boson production processes have production cross-sections around the femtobarn.
Despite this, several tri-boson production processes have been measured at the LHC by the
ATLAS and CMS collaborations~\cite{CMS:2020hjs, ATLAS:2015ify, ATLAS:2016jeu, ATLAS:2017bon}.
These processes are particularly interesting for BSM searches due to their 
sensitivity to anomalous gauge couplings and potential new physics in the 
electroweak sector, see Fig. \ref{fig:tri-boson-diagrams}\footnote{The diagrams are made using FeynGame \cite{harlander2020feyngame, harlander2024feyngame}}.
This analysis have been made possible by the recent 
implementation of the BLHA interface
in \Vbfnlo{} 3.0, which is also available for the rest of the tri-boson
and di-boson processes with fully leptonic final states.

We place particular emphasis on quantifying theoretical uncertainties
from multiple sources. These uncertainties arise from two main factors:
the truncation of the perturbative series in computing the hard 
collision matrix element, and the arbitrary cut-off scale in parton 
shower evolution that delineates the boundary between perturbative and
non-perturbative scales. Understanding these uncertainties is crucial 
for interpreting experimental results and constraining BSM physics. We 
estimate them through variations of the factorization, renormalization, 
and hard shower scales.

Additionally, we study uncertainties related to shower-induced migration 
effects across kinematic cuts. The inclusion of generation-level cuts affects
predictions at both the integrated and qS differential cross-section levels, as 
events that could potentially migrate into the accepted region after showering
may be excluded at generation. This effect is particularly significant for 
safety cuts that are required for the simulation and influence QCD radiation, 
such as the Frixione Isolation cut. Furthermore, determining whether such cuts 
make predictions more or less inclusive requires dedicated analysis and cannot
be assumed a priori. The resulting migration effects represent a 
source of uncertainty in the final prediction that must be carefully quantified.

This work is structured as follows: In Sec. \ref{sec:cal}, we give the
calculational setup, including programs, cuts, and other input parameters
necessary to reproduce the results presented. In Sec. \ref{sec:pheno}, the
differential distributions of different observables, with their corresponding
scale variations at LO+PS, NLO and NLO+PS are presented and
we study the migration effects of some generation level cuts. 
In Sec. \ref{sec:conclu}, we present our conclusions.

\section{Calculational Setup}\label{sec:cal}

For this study, we use \Herwig 7.3 as the Monte Carlo event generator and \Vbfnlo{} 3.0
as the NLO QCD amplitude provider.
The communication between these programs is facilitated through the BLHA interface.

\subsection{NLO+PS Matching using \Herwig 7.3}
The \Herwig 7.3 Monte Carlo event generator
enables NLO event simulation through extensions to the previously developed
Matchbox module~\cite{Platzer:2011bc}, allowing external amplitude providers to
evaluate tree-level and one-loop matrix elements. These elements are
automatically combined with the Catani-Seymour dipole
subtraction~\cite{Catani:1996vz}, and general-purpose and specialized phase
space generation algorithms, resulting in a full NLO calculation. This NLO
calculation can be further extended by automatically determined matching
subtractions, which combine it with a downstream parton shower algorithm.

While the Matchbox module supports a range of dedicated hard process
calculations, communication with external general-purpose amplitude providers,
such as \Vbfnlo{} 3.0, is
facilitated through extensions to the BLHA 2 standard~\cite{Binoth:2010xt,
  Alioli:2013nda}.

\subsection{VBFNLO 3.0}
\Vbfnlo{} 3.0 is a flexible Monte
Carlo generator at the parton level for processes involving electroweak bosons.
It is capable of providing a fully differential distributions at fixed-order
parton level NLO QCD accuracy. Communication between \Vbfnlo{} 3.0 and \Herwig 7.3 occurs via
an interface built on the BLHA 2 standard~\cite{Binoth:2010xt, Alioli:2013nda}.
\Vbfnlo{} 3.0 has been extended to allow access to additional features such as the
internal phase-space generator, which is used in the results presented in this
article. However, these extensions are optional, and a Monte Carlo generator
that complies with standard protocols is sufficient for obtaining amplitudes
from \Vbfnlo{} 3.0.

The interface for Vector Boson Scattering of single and double vector boson
production was previously performed and detailed results for the $pp \to WWjj$
production channel were discussed in Ref.~\cite{Rauch:2016upa}. In this
article, we extend the interface to all the double and triple boson production
processes with fully leptonic decays and present detailed results for the process:
\begin{equation}
  \label{eq:process}
  pp \to e^{+} \nu_e \mu^{-} \bar{\nu}_\mu \gamma   \qquad  \qquad (W ^+W^-\gamma)
\end{equation}
The calculation is performed at order $\alpha_s \alpha^2$ considering
the massless fermion limit. The leptonic decay of the $W$-vectors are
fully included, considering all off-shell effects and spin correlations.
In \fig{fig:tri-boson-diagrams}, some representative graphs at LO showing
the rich structure of the process can be found, including diagrams
sensitive to quartic (top-left) and triple (bottom-left) gauge-boson
verticies. It also shows examples of photons radiated off the leptons
(top-right) or the quarkline (bottom-right) graphs which are considered
to be irreducible backgrounds in the context of anomalous gauge coupling searches.
The calculation of this process was presented first in Ref.~\cite{Bozzi:2010sj}
and release as part of the package \Vbfnlo{} 2.7\cite{Arnold:2008rz}.
We refer the reader to those references for details on the calculations
and the test performed to cross-checked the calculation.

The accuracy of NLO fixed order predictions generated by the \Herwig 7.3 +\Vbfnlo{} 3.0
through the BHLA 2 interface setup has undergone extensive validation against
standalone calculations obtained from \Vbfnlo{} 3.0 predictions, using both a range
of integrator and phase space generation algorithms supplied by the standard
Matchbox modules or by utilizing the versatile interface structure to access
the appropriate \Vbfnlo{} 3.0 routines.

\subsection{Matching algorithms and Uncertainties}
\label{sec:match}
The Matchbox module provides a wide range of options for generating matched cross sections via 
various algorithms, including direct and subtractive matching to both angular-ordered and dipole showers, 
as well as multiplicative (Powheg type) matching. Matchbox leverages adaptive methods to sample 
matrix-element corrections and adds, in case of angular-ordered showers,  
truncated showers to account for large-angle soft and hard collinear
emissions on top of it.

To estimate theoretical uncertainties, we employ a combination of
scale variations for the hard process and the parton shower.
For the hard process, we use the 7-point variation scheme for 
the factorization $(\mu_F)$ and renormalization $(\mu_R)$ scales. This 
involves varying these scales independently by factors of 0.5
and 2 around the central scale, excluding the extreme variations
$(0.5, 2)$ and $(2, 0.5)$.
Independently, we vary the hard veto scale $(\mu_Q)$ of the shower
by factors of 0.5 and 2. This approach results in a total of 9
variations: 7 from the hard process scale variations and 2 from 
the shower scale variations.
We choose this method under the assumption that the shower and
hard process variations are largely independent. A full 
combination of all scale variations would require significantly 
more computational resources. This approach provides a reasonable 
estimate of the uncertainties while remaining computationally 
feasible.

The \texttt{resummation} profile scale is
used for both showers and the matrix-element correction entering the
multiplicative matching to ensure a smooth transition between the hard matching
and resummation regions, while maintaining the resummation properties of the
parton shower.

\subsection{Cuts and phase-space optimization}

In this subsection, we detail the parameters and settings used
in our parton-level study of the process
$pp \rightarrow e^{+}\nu_e \mu^{-}\bar{\nu}_{\mu}\gamma + X$
at $\sqrt{s} = 13.6\, \rm TeV$.

\subsubsection{General Settings}
We adopt a massless approximation for all partons and do not include
multiple parton interactions (MPI). The \Herwig 7.3 shower modules are used
with their default settings.

\subsubsection{PDF and EW parameters}
We use the CT18 PDF with four active flavours. The EW
parameters are set to the standard values used in \Herwig 7.3:
\begin{align}
  m_W & = 80.3770\; {\rm GeV}, \; \Gamma_W = 2.085\; {\rm GeV} \nonumber         \\
  m_Z & = 91.1876\; {\rm GeV}, \; \Gamma_Z = 2.495\; {\rm GeV} \nonumber        \\
  G_F & = 1.16637\cdot10^{-5}\; {\rm GeV^{-2}}.
\end{align}

\subsubsection{Scale Choices}
For both factorization and
renormalization scales, we use the mass of the full EW system.

\subsubsection{Kinematic Cuts and Phase Space Optimization}
We implement two sets of cuts: generation-level and analysis-level. Our 
default generation-level cuts, which we call ``inclusive phase-space", are:
\begin{align}\label{eq:inclusive-phase-space}
   & p_{T, \ell} > 10\, {\rm GeV},\quad |\eta_{\ell}| < 5.0 \nonumber     \\
   & p_{T, \gamma} > 10\, {\rm GeV}, \quad|\eta_{\gamma}| < 5.0, \nonumber \\
   & \Delta R_{\ell \gamma} > 0.3,\quad                                    
\end{align}
The analysis-level cuts are tighter:
\begin{align}\label{eq:analysis:cuts}
   & p_{T, \ell} > 30\, {\rm GeV},\quad |\eta_{\ell}| < 2.5 \nonumber  \\
   & p_{T, \gamma} > 30\, {\rm GeV}, \quad|\eta_{\gamma}| < 2.5,\nonumber \\
   & \Delta R_{\ell \gamma} > 0.7,\quad   
\end{align}

For photon isolation, we use the Frixione algorithm with $\delta = 0.4$,
$\epsilon = 0.05$. Jets are defined using the anti-kt algorithms
with $R=0.4$ and $p_T > 20\, {\rm GeV}$.

The runs have been made using an pre-optimized phase-space grid,
which we refer to as the ``tight phase-space", applying the following cuts:
\begin{align}\label{eq:tight-phase-space}
& p_{T, \ell} > 25, {\rm GeV},\quad |\eta_{\ell}| < 3.0 \nonumber  \\
& p_{T, \gamma} > 25, {\rm GeV}, \quad|\eta_{\gamma}| < 3.0,\nonumber \\
& \Delta R_{\ell \gamma} > 0.5.
\end{align}

The primary purpose of this optimization is to enhance event generation
efficiency while preserving the physical integrity of the predictions.
By setting optimization cuts slightly looser than the analysis cuts, 
we ensure adequate sampling of events near the analysis cut boundaries.

The grid optimization and phase-space integration is performed using the MONACO sampler,
a customized implementation of the VEGAS algorithm. Importantly,
even when specific kinematic cut values are set during the grid pre-optimization phase-space  
integration, the event generation still produces events outside 
these ranges, provided they satisfy the generation-level cuts. This
behavior occurs because MONACO optimizes a rectangular grid that 
increases point density in regions with large cross-sections, while
still sampling points with wider grid spacing in regions with low cross-sections.

For clarity, the cuts in Eq. \ref{eq:tight-phase-space} are used solely to generate an
optimized grid, improving the convergence of event generation. We maintain the cuts specified in Eq. \ref{eq:inclusive-phase-space}
and Eq. \ref{eq:analysis:cuts} throughout all runs.

\begin{figure}
  \centering
  \includegraphics[width=0.46\textwidth]{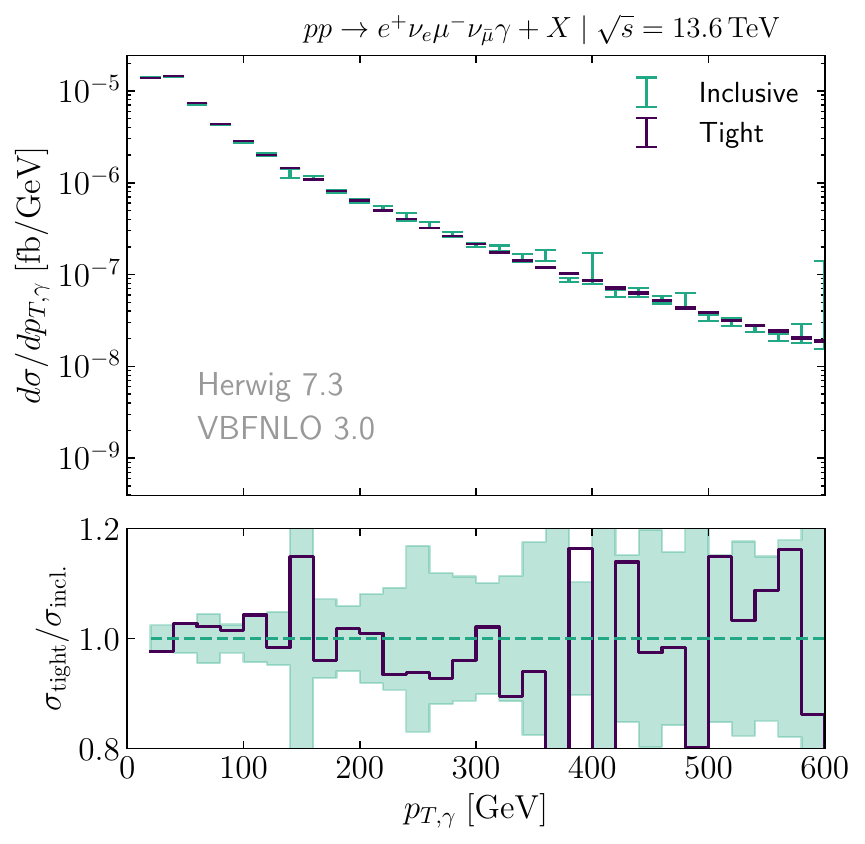}
  \caption{Comparison between integrated cross-sections calculated using tight
  and inclusive phase space grids.
  Top panel: The differential cross-section $d\sigma /dp{T, \gamma}$ [fb/GeV] as a function of photon transverse momentum $p{T, \gamma}$ [GeV]. Purple curve shows results with tight phase space grid, blue curve shows inclusive phase space grid results, with statistical uncertainties indicated by error bars. Bottom panel: Ratio of tight to inclusive phase space cross-sections, with the shaded region representing the combined statistical uncertainty propagated from both calculations.}
  \label{fig:phase-space-check}
\end{figure}

We have verified that this optimization introduces no artifacts by comparing differential
distributions with lower-statistics runs using the generation level cuts to optimize the phase-space grid,
Eq. \ref{eq:inclusive-phase-space}.
Fig. \ref{fig:phase-space-check} demonstrates this comparison for the photon transverse
momentum distribution. The excellent agreement between the tight and inclusive phase-space
results confirms the validity of our approach, ensuring that our predictions remain physically
meaningful while benefiting from improved computational efficiency.

\section{Phenomenological results}
\label{sec:pheno}

In this section, we present and analyze the results of our parton shower matching study
for the tri-boson production process $pp \rightarrow e^{+}\nu_e \mu^{-}\bar{\nu}_{\mu}\gamma + X$
at $\sqrt{s} = 13.6\, \rm TeV$.
We focus on two key aspects: scale variations and migration effects. 

Scale variations provide crucial insights into the theoretical uncertainties arising from the
truncation of the perturbative series and the arbitrary separation between perturbative and non-perturbative
regimes. By examining these variations, we can assess the reliability of our 
predictions and identify regions where higher-order corrections may be significant.

Migration effects, on the other hand, help us understand how parton shower emissions can shift
event kinematics across cut boundaries, potentially impacting experimental analyses.

Our analysis encompasses fixed-order NLO results, LO+PS predictions, and NLO+PS simulations using 
both dipole and angular-ordered showers. By comparing these different approaches, we aim to 
elucidate the impact of parton showers on various observables and provide guidance for future 
experimental and theoretical studies of multi-boson production processes.

\subsection{Scale variations}
\label{subsec:scale_var}
\begin{figure}
  \centering
  \includegraphics[width=0.46\textwidth]{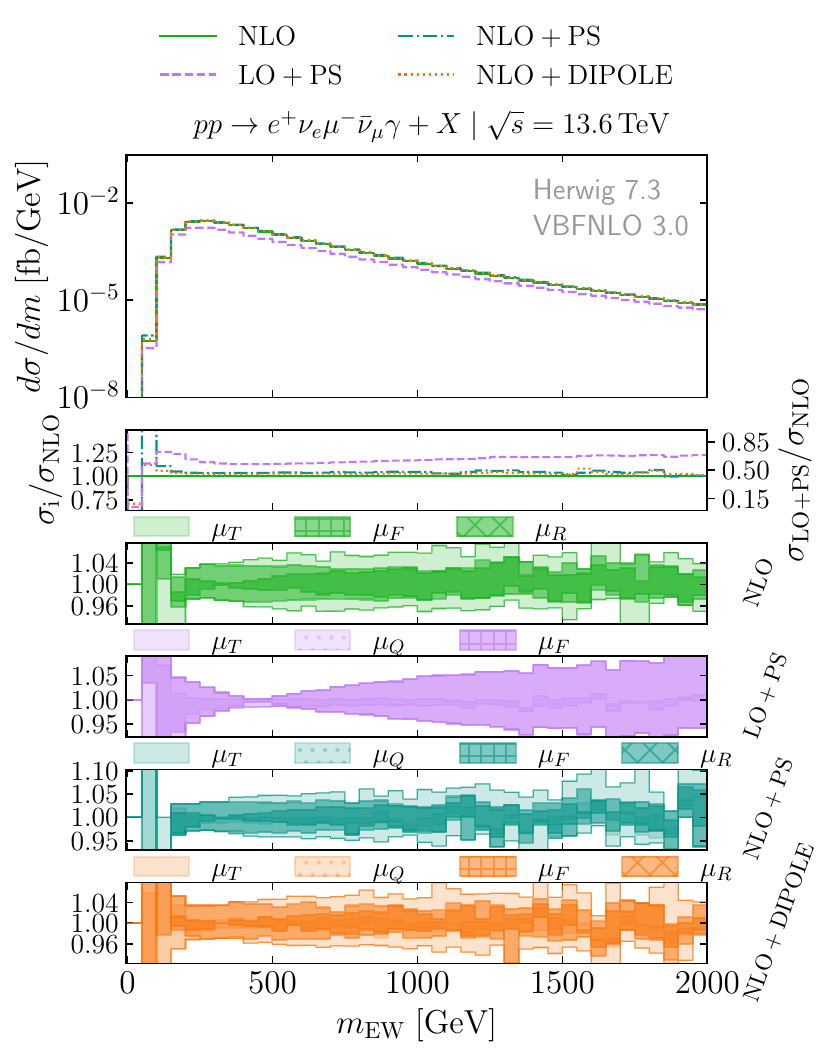}
  \caption{Distribution of the electroweak system invariant mass showing fixed-order 
  NLO (green), LO+PS (pink), and NLO matched predictions with angular-ordered (blue)
  and dipole (orange) showers. Top ratio panel compares all predictions to the NLO 
  fixed-order result, with the LO+PS ratio shown on a separate scale (right axis). 
  Lower panels display scale variation bands for each prediction type, with increasing
  opacity indicating the cumulative effect of different scale variations: factorization
  and renormalization scales (darkest), hard veto scale, and total combined 
  uncertainty (lightest).}
  \label{fig:scale_EWSystemMass}
\end{figure}
\begin{figure}
  \centering
  \includegraphics[width=0.46\textwidth]{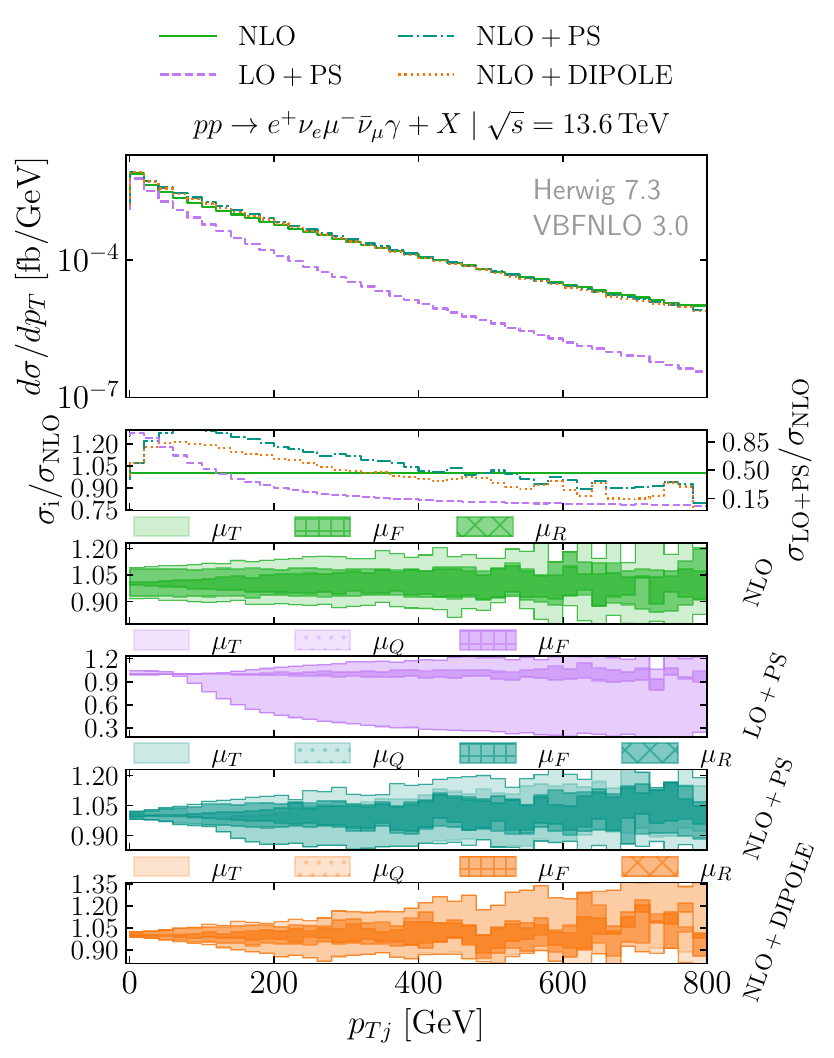}
  \caption{Distribution of the leading jet transverse momentum showing fixed-order 
  NLO (green), LO+PS (pink), and NLO matched predictions with angular-ordered (blue)
  and dipole (orange) showers. Top ratio panel compares all predictions to the NLO 
  fixed-order result, with the LO+PS ratio shown on a separate scale (right axis). 
  Lower panels display scale variation bands for each prediction type, with increasing
  opacity indicating the cumulative effect of different scale variations: factorization
  and renormalization scales (darkest), hard veto scale, and total combined 
  uncertainty (lightest).}

  \label{fig:scale_ptj1}
\end{figure}

Fig. \ref{fig:scale_EWSystemMass} presents a comprehensive analysis of the 
electroweak system's invariant mass distribution and its associated scale 
uncertainties. The main panel shows differential distributions calculated 
using various approaches: fixed-order NLO calculations (green), 
LO+PS predictions (pink), and NLO-matched results using both dipole (orange) 
and angular-ordered (blue) showers.

The uppermost ratio panel compares each prediction to the fixed-order NLO result,
with the LO+PS ratio displayed on a separate scale (right axis) due to its larger variation.
The four lower panels illustrate the cumulative impact of different scale variations on
the differential cross-section. These panels use increasing opacity to show
the sequential effects of varying the factorization ($\mu_F$) and 
renormalization ($\mu_R$) scales (darkest band), the hard veto scale ($\mu_Q$),
and finally, the total combined uncertainty ($\mu_T$, lightest band).
All scale variations follow the scheme detailed in Sec. \ref{sec:match}.

The invariant mass distribution of the EW-system is minimally affected by parton showering 
at NLO, and both, the angular-ordered and dipole showers, produce similar predictions for 
the central values and scale variations. The uncertainty in the scale variation is dominated by the 
renormalization and factorization scales in this observable, with a total scale variation of 
$\pm5\%$. Notably, the LO+PS central value fails to predict the NLO result both in shape 
and normalization. This discrepancy arises from substantial NLO contributions, due to the
colorless nature of the LO production process final state, including significant new contributions from
gluon-initiated processes and the emergence of new phase-space regions. These effects 
manifest as a hard jet recoiling against the electroweak system, 
highlighting the importance of full NLO-matched results.

In contrast, jet observables are markedly influenced by parton shower effects, as illustrated
in Fig. \ref{fig:scale_ptj1}, which shows the differential distribution of the leading jet's
transverse momentum. At LO+PS, the jet consists solely of parton shower radiation, while in 
the NLO matched simulation, it is present at the matrix element level. Consequently,
we anticipate larger uncertainties and a more pronounced impact of showering. 
In the NLO+PS scenario, the jet's transverse momentum distribution remains relatively
stable with respect to shower effects at low to moderate $p_T$, typically within 
$10-20\%$ of the fixed-order prediction.

The scale variation uncertainties exhibit a particularly dramatic improvement when moving from
LO+PS to NLO+PS predictions. At LO+PS, variations are dominated by the hard veto scale,
reaching up to $70\%$ in the tail of the distribution. The inclusion of real radiation
at NLO substantially reduces these uncertainties, 
resulting in more moderate variations ranging from a few percent at low $p_T$
to approximately $\pm20\%$ at $p_{T,j} = 800\; \rm GeV$.

\subsection{Study of the migration effects}
\label{subsec:migration}

In this subsection, we examine the impact of migration through cut boundaries resulting from 
kinematical shifts induced by additional radiation in the parton shower.
This phenomenon is crucial for understanding the interplay between theoretical
calculations and experimental measurements.

Selection cuts in experimental analyses should ideally be applied at the event
level after the parton shower process while keeping the fixed order prediction
fully inclusive. However, this approach can significantly
reduce computational efficiency, particularly for processes with highly exclusive
cuts such as Vector Boson Scattering (VBS), since many generated events will be
discarded by the analysis cuts.

In some cases, generation-level cuts cannot be avoided, such as when safety cuts 
are required to prevent theoretical divergences. Examples include the Frixione 
Isolation cut or minimum separation requirements between photons and charged leptons.
Moreover, when events can have negative weights, the impact of these cuts becomes
highly non-trivial – modifying cut parameters may either increase or decrease the
total cross-section in ways that cannot be predicted without explicit calculation.
This makes it impossible to determine a priori whether certain cut choices lead to
more inclusive or exclusive selections.

In these cases, the effect of generation-level cuts must be incorporated as an additional 
theoretical uncertainty in the prediction. This uncertainty can often be estimated by
varying the parameters of the generation-level cuts.

The effects of the parton shower in relation to cuts can be categorized into two distinct
phenomena. Firstly, discrepancies between parton shower and fixed-order kinematics lead to 
variations in differential distributions due to analysis-level cuts. Secondly, the presence of 
generation-level cuts can induce migration effects, whereby shower-induced kinematic changes may 
allow events initially excluded by generation cuts to potentially fall within the analysis-permitted
kinematic region. 

We now focus on the migration effects as defined previously. Our analysis examines the impact
of modifying the generation-level cuts for charged leptons and photon selection criteria.
We present two types of results: the effect on the integrated cross-section and on
differential distributions.

\begin{figure}
  \centering
  \includegraphics[width=0.46\textwidth]{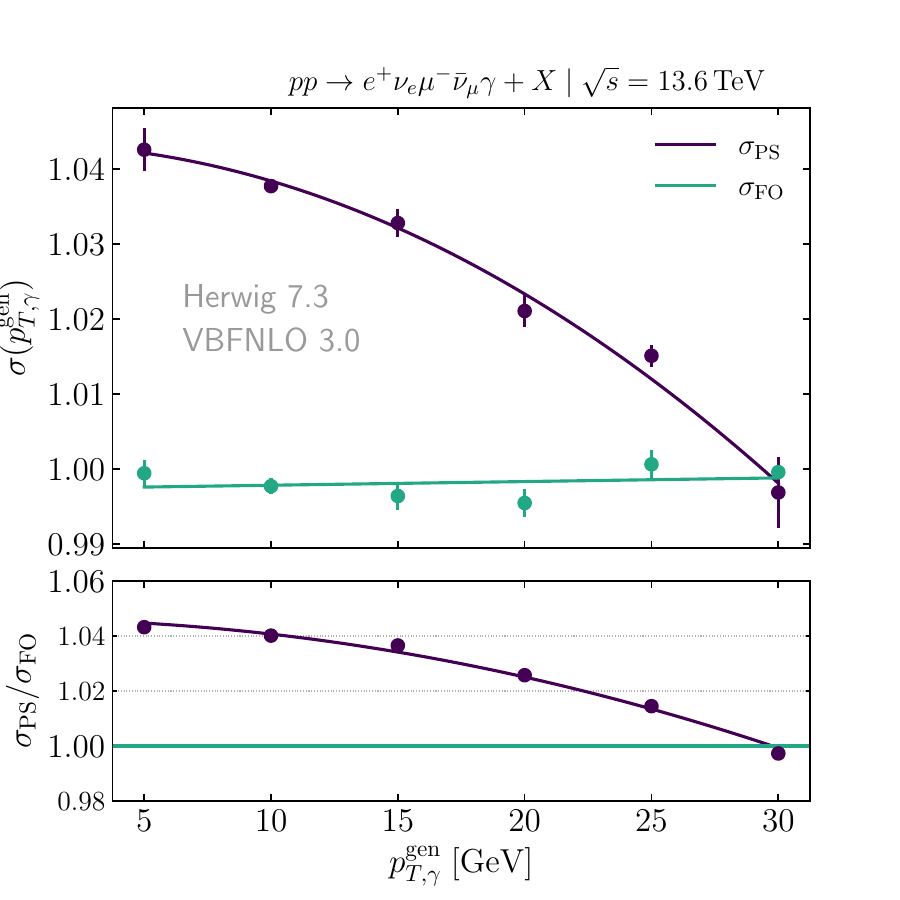}
  \caption{Dependence of the integrated cross-section of the generation-level
  photon $p_{T, \gamma}^{\rm gen}$ cut.
  Top panel: integrated cross-section $\sigma(p_{T, \gamma}^{\rm gen})$ for 
  parton shower (PS) and fixed-order (FO).
  Bottom panel: Ratio of the PS over FO cross-section.}
  \label{fig:photon-ptcut-int-xsec}
\end{figure}

Fig. \ref{fig:photon-ptcut-int-xsec} illustrates the impact of varying the photon
transverse momentum generation threshold on the integrated cross-section. The purple
curve represents the parton shower result, while the blue curve shows the fixed-order
calculation. As anticipated, the fixed-order calculation remains largely unaffected by the 
generation-level cut, with minor variations attributable to statistical fluctuations.
This behavior is expected because fixed-order calculations do not include the additional
radiation that can cause migration effects.

In contrast, the parton shower exhibits a more pronounced effect, 
with the integrated cross-section decreasing as the generation-level cut approaches
the analysis cut. This trend can be understood as follows: as the generation-level cut
approaches the analysis cut, there's less phase space for events to migrate into the
analysis region due to shower effects, leading to a decrease in the accepted cross-section.
This observation underscores the importance of careful consideration of generation-level
cuts in shower-matched predictions, particularly for processes with tight kinematic selections.

\begin{figure}
  \centering
  \includegraphics[width=0.46\textwidth]{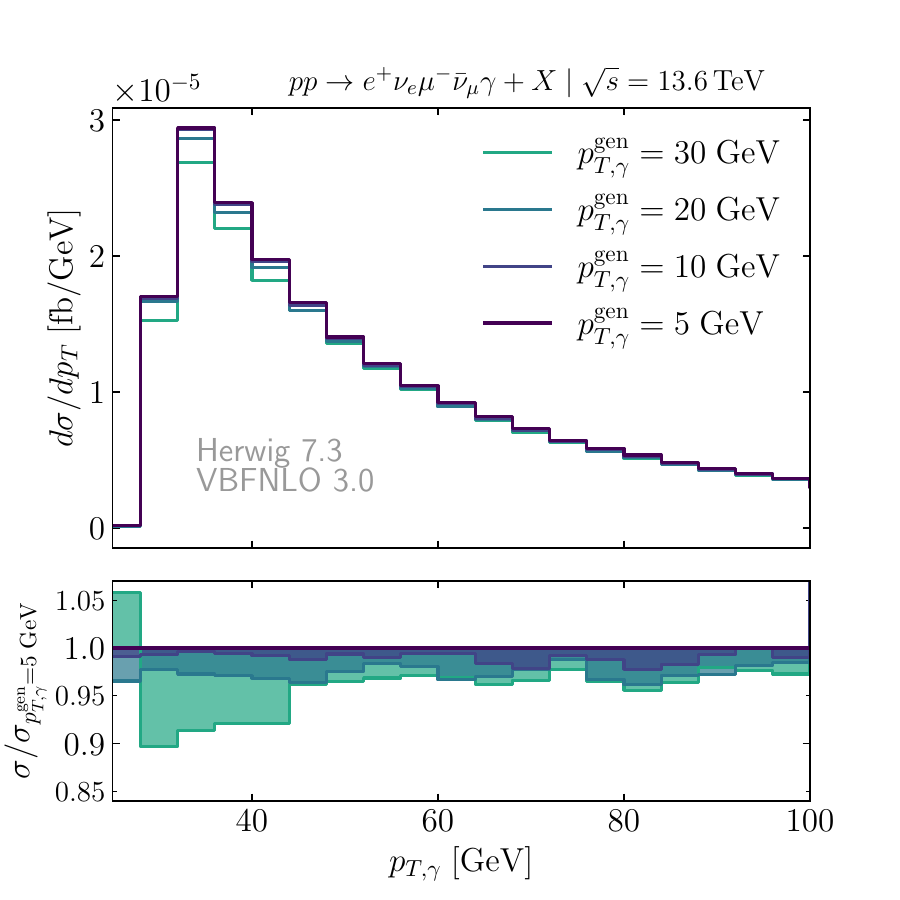}
  \caption{Differential cross-section for the process $p p \rightarrow
  e^{+}\nu_e \mu^{-} \bar{\nu}_{\mu} \gamma + X$ at 
  $\sqrt{s} = 13.6\, {\rm TeV}$.
  Top panel: The differential cross-section $d\sigma /dp_{T, \gamma}$ [fb/GeV]
  is shown as a
  function of the photon transverse momentum $p_{T, \gamma}$ [GeV]. Four curves 
  for different generation-level photon $p_{T, \gamma}^{\rm gen}$ cuts:
  5 GeV, 10 GeV, 20 GeV, and 30 GeV.
  Bottom panel: Ratio of the cross-sections for different
  $p_{T, \gamma}^{\rm gen}$ generation cuts relative to the 5 GeV generation cut.}
  \label{fig:photon-ptcut-diff-xsec}
\end{figure}

To elucidate the impact on differential distributions, 
we refer to Fig. \ref{fig:photon-ptcut-diff-xsec}. This figure displays the
differential distribution $d\sigma/dp_{T,\gamma}$ as a function of the photon 
transverse momentum $p_{T,\gamma}$. The main plot features four curves corresponding
to different values of the generation-level cut: $5\, \rm GeV$, 
$10\, \rm GeV$, $20\, \rm GeV$, $30\, \rm GeV$. The lower panel presents the ratio 
with respect to the lowest value of the transverse momentum generation-level cut 
($5\, \rm GeV$), with the filled region indicating the missing cross-section.

Examining the figure reveals that the differential distribution 
is not uniformly affected across all momentum ranges.
The effects are more pronounced at lower transverse momenta, 
specifically for $p_{T,\gamma} < 50 \, \rm GeV$. This observation can be attributed to
the proximity to the cut threshold, where smaller kinematic shifts induced by the
shower are sufficient to exclude events from the analysis.
As $p_{T,\gamma}$ increases, the curves for different generation-level cuts
converge, indicating that migration effects become less significant
for high-$p_T$ photons, where the shower-induced kinematic shifts are relatively
small compared to the photon's transverse momentum.

\begin{figure}
  \centering
  \includegraphics[width=0.46\textwidth]{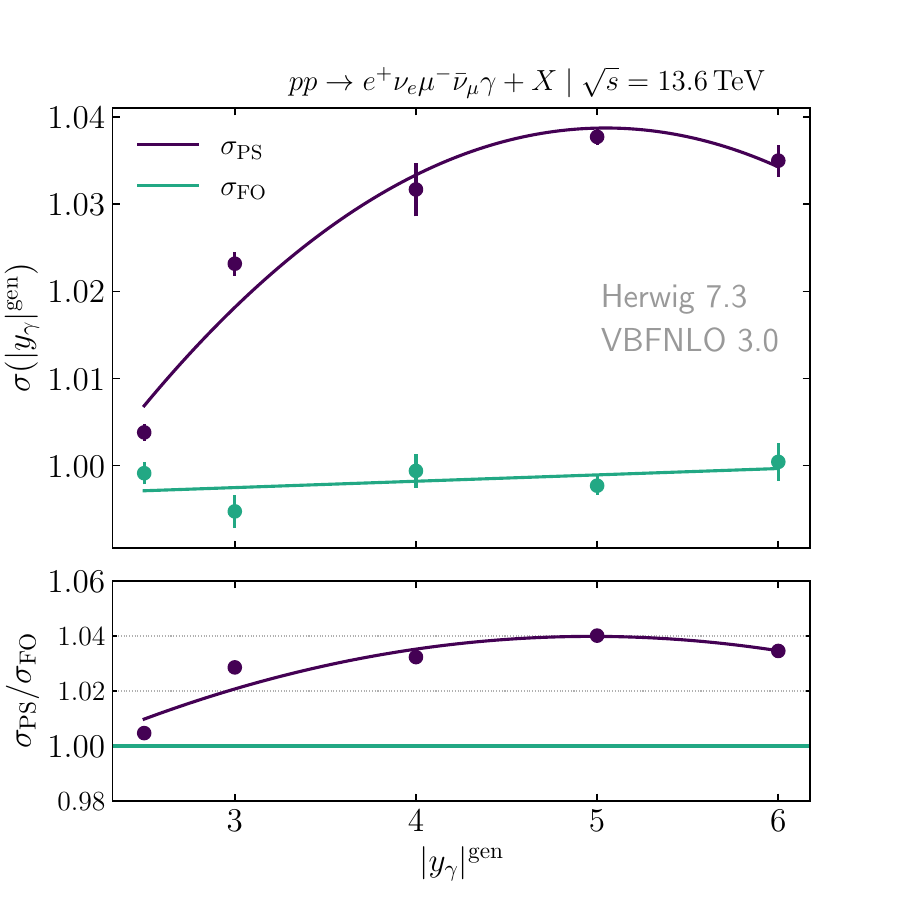}
  \caption{Dependence of the integrated cross-section of the generation-level
  photon $|y_\gamma|^{\rm gen}$ cut.
  Top panel: integrated cross-section $\sigma(|y_\gamma|^{\rm gen})$ for 
  parton shower (PS) and fixed-order (FO).
  Bottom panel: Ratio of the PS over FO cross-section.}
  \label{fig:photon-ycut-int-xsec}
\end{figure}

\begin{figure}
  \centering
  \includegraphics[width=0.46\textwidth]{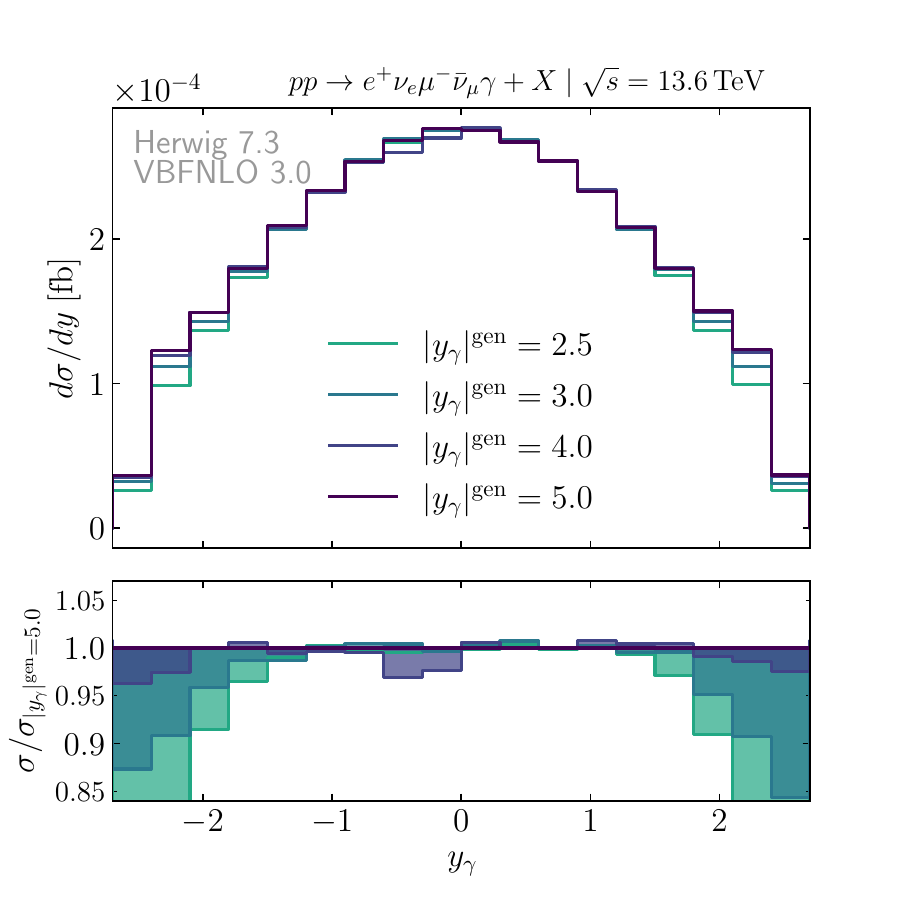}
  \caption{Differential cross-section for the process $p p \rightarrow
  e^{+}\nu_e \mu^{-} \bar{\nu}_{\mu} \gamma + X$ at 
  $\sqrt{s} = 13.6\, {\rm TeV}$.
  Top panel: The differential cross-section $d\sigma /dy_{\gamma}$ [fb]
  is shown as a
  function of the photon rapidity $y_{\gamma}$. Four curves 
  for different generation-level photon $y$ cuts 
  ($|y_{\gamma}|^{\rm gen}$): 2.5, 3.0, 4.0 and 5.0.
  Bottom panel: Ratio of the cross-sections for different
  $|y_{\gamma}|^{\rm gen}$ generation cuts relative to the 5.0 generation cut.}
  \label{fig:photon-ycut-diff-xsec}
\end{figure}

\begin{figure}
  \centering
  \includegraphics[width=0.46\textwidth]{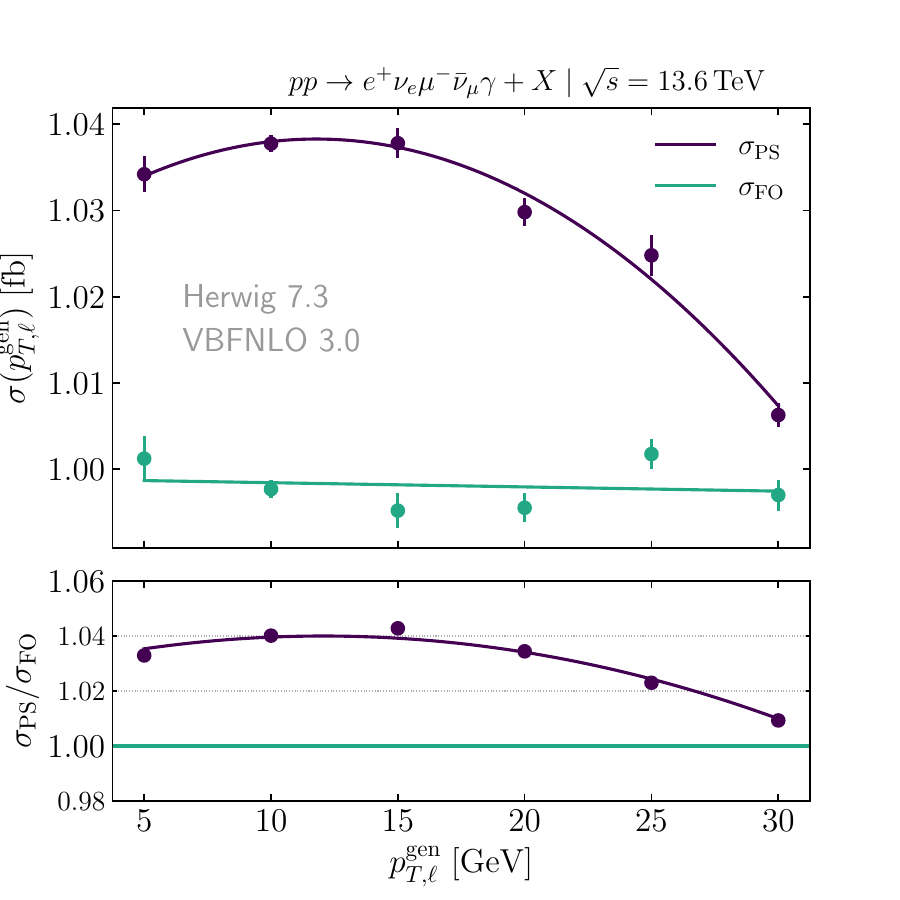}
  \caption{Dependence of the integrated cross-section of the generation-level
  charged lepton $p_{T, \ell}^{\rm gen}$  cut.
  Top panel: integrated cross-section $\sigma(p_{T, \ell}^{\rm gen})$ for 
  parton shower (PS) and fixed-order (FO).
  Bottom panel: Ratio of the PS over FO cross-section.}
  \label{fig:lepton-ptcut-int-xsec}
\end{figure}

\begin{figure}
  \centering
  \includegraphics[width=0.46\textwidth]{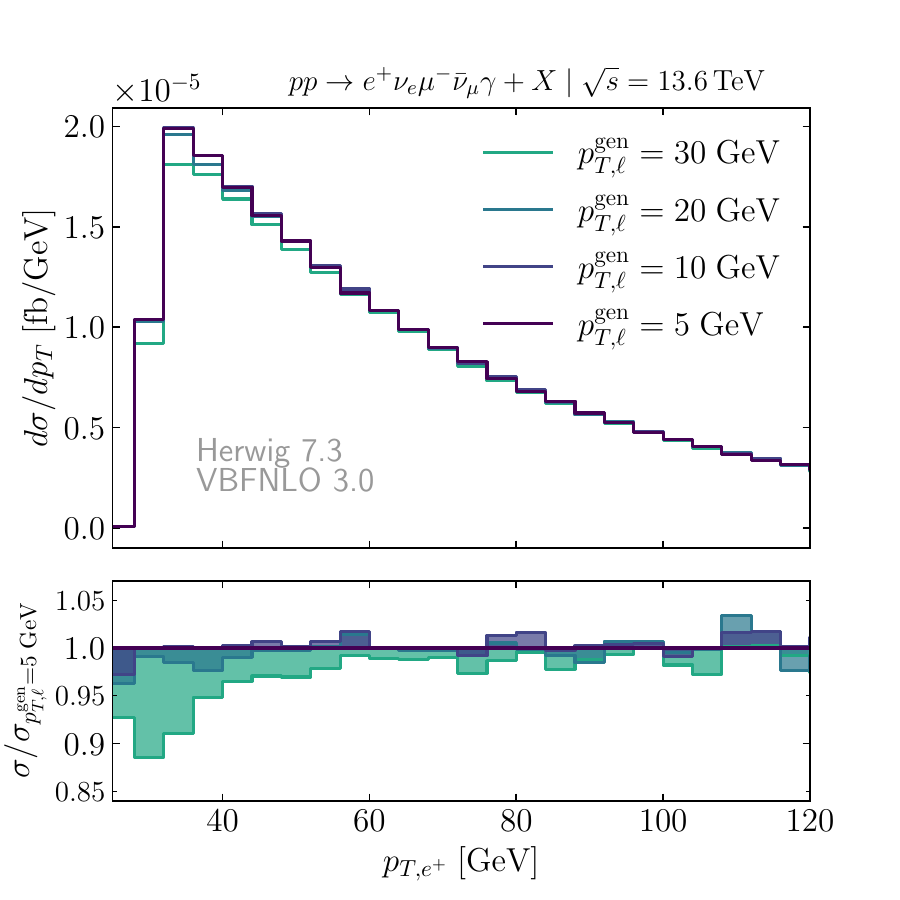}
  \caption{Differential cross-section for the process $p p \rightarrow
  e^{+}\nu_e \mu^{-} \bar{\nu}_{\mu} \gamma + X$ at 
  $\sqrt{s} = 13.6\, {\rm TeV}$.
  Top panel: The differential cross-section $d\sigma /dp_{T, e^{+}}$ [fb/GeV]
  is shown as a
  function of the positron transverse momentum $p_{T, e^{+}}$ [GeV]. Four curves 
  for different generation-level charged lepton $p_{T, \ell}^{\rm gen}$
  cuts: 5 GeV, 10 GeV, 20 GeV, and 30 GeV.
  Bottom panel: Ratio of the cross-sections for different
  $p_{T, \ell}^{\rm gen}$ generation cuts relative to the 
  5 GeV generation cuts.}
  \label{fig:lepton-ptcut-diff-xsec}
\end{figure}

\begin{figure}
  \centering
  \includegraphics[width=0.46\textwidth]{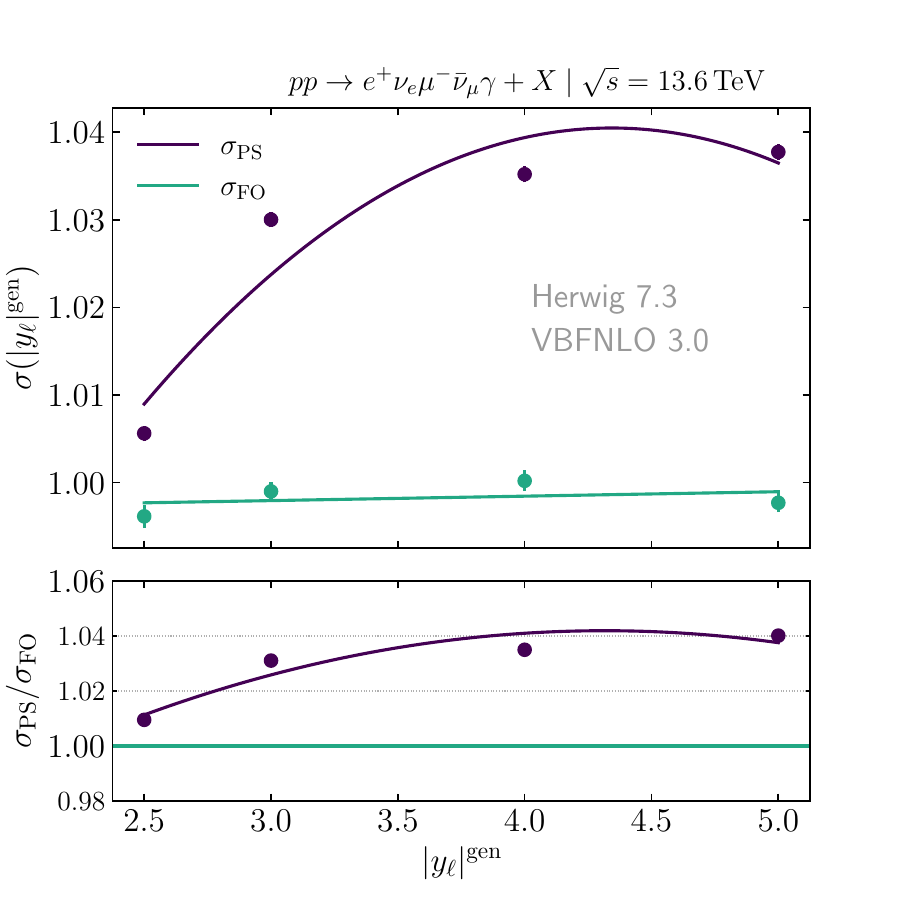}
  \caption{Dependence of the integrated cross-section of the generation-level
  charged lepton $|y_{\ell}|^{\rm gen}$ cut.
  Top panel: integrated cross-section $\sigma(|y_{\ell}|^{\rm gen})$ for 
  parton shower (PS) and fixed-order (FO).
  Bottom panel: Ratio of the PS over FO cross-section.}
  \label{fig:lepton-ycut-int-xsec}
\end{figure}

\begin{figure}
  \centering
  \includegraphics[width=0.46\textwidth]{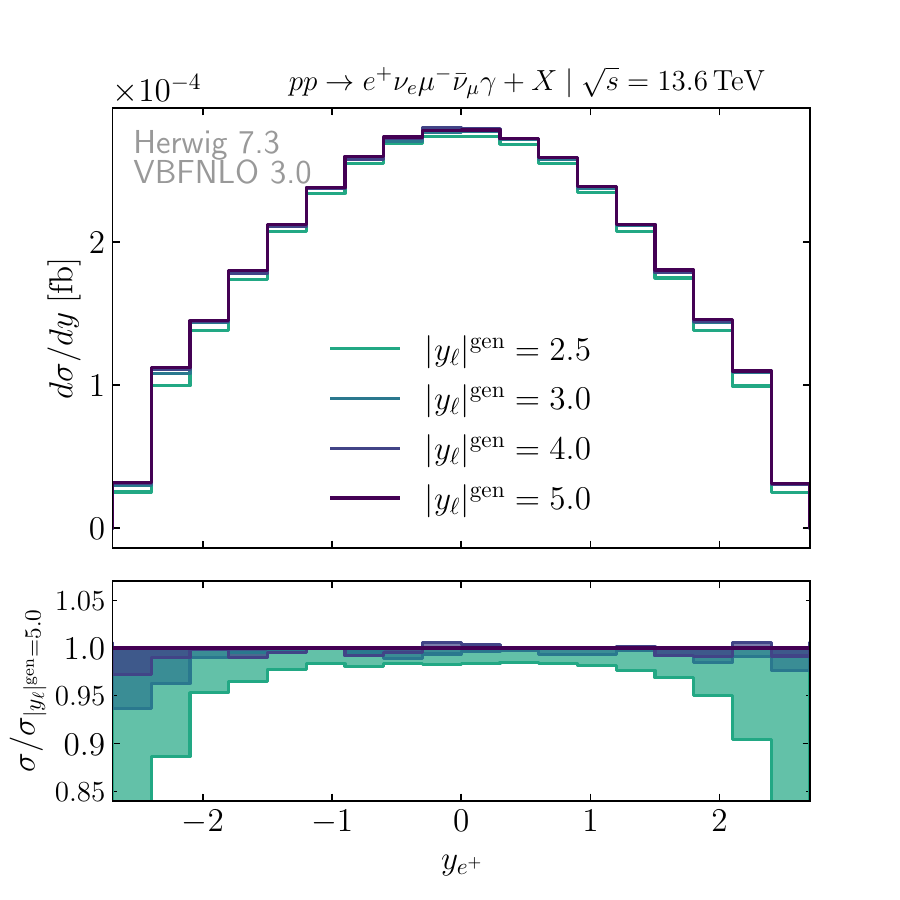}
  \caption{Differential cross-section for the process $p p \rightarrow
  e^{+}\nu_e \mu^{-} \bar{\nu}_{\mu} \gamma + X$ at 
  $\sqrt{s} = 13.6\, {\rm TeV}$.
  Top panel: The differential cross-section $d\sigma /dp_{T, e^{+}}$ [fb/GeV]
  is shown as a
  function of the positron rapidity $y_{e^{+}}$ [GeV]. Four curves 
  for different generation-level charged lepton $|y_{\ell}|^{\rm gen}$ cuts:
  2.5, 3.0, 4.0 and 5.0.
  Bottom panel: Ratio of the cross-sections for different
  $|y_{\ell}|^{\rm gen}$ generation cuts relative to the 5.0 generation cut.}
  \label{fig:lepton-ycut-diff-xsec}
\end{figure}

Fig. \ref{fig:photon-ycut-int-xsec} presents the resulting 
integrated cross-sections as a function of the photon's rapidity generation-level
cut, $\sigma(|y_{\gamma}|^{\rm gen})$, where we 
observe patterns similar to those in the transverse momentum study. The fixed-order 
prediction (blue curve) maintains its expected insensitivity to the rapidity cut, while
the parton-showered calculation (purple curve) exhibits a systematic decrease in 
cross-section as the rapidity cut becomes more restrictive 
(i.e., with decreasing $|y_{\gamma}|^{\rm gen}$).
The corresponding differential distribution, shown in Fig. \ref{fig:photon-ycut-diff-xsec},
reveals that the 
most pronounced effects occur near the cut boundary, with variations exceeding $10\%$
at $|y_{\gamma}|^{\rm gen} = 3.0$. 
Figs. \ref{fig:lepton-ptcut-int-xsec}-\ref{fig:lepton-ycut-diff-xsec} present analogous results for 
the positron, showing both the integrated cross-sections and differential distributions. 
These results mirror those observed for the photon and are included for completeness
without further discussion.

Although our analysis uses more extreme generation-level cuts than those typically
employed in experimental settings, we observe that these cuts can impact
differential distributions by up to $10\%$. This effect becomes particularly significant 
when compared to scale variation uncertainties of approximately $\pm5\%$. The implications 
are especially relevant for analyses using highly inclusive cuts or processes requiring 
stringent kinematic constraints, such as Vector Boson Scattering (VBS). Furthermore, 
certain cuts like the Frixione Isolation cannot be removed, and estimating the parton shower effect and their associated uncertainties due to this cut
is non-trivial. A more comprehensive treatment of such cases would require the inclusion
of fragmentation functions to eliminate the need for these cuts altogether.

These findings underscore the critical interplay between parton shower effects and 
kinematic cuts, highlighting the necessity for careful consideration of generation-level 
constraints in precision phenomenological studies of multi-boson production processes. 
This is particularly important for ensuring reliable theoretical predictions that
can be meaningfully compared with experimental measurements.

\section{Conclusions}
\label{sec:conclu}

\Vbfnlo{} 3.0 now incorporates BLHA interface support for all di-boson
and tri-boson processes with fully leptonic final states, substantially
enhancing our capabilities to study multi-boson production processes.
This implementation enables NLO+PS matched calculations across a wide range
of multi-boson channels, facilitating precise predictions that combine 
the accuracy of fixed-order calculations with the detailed final-state 
modeling of parton showers.

Our comprehensive analysis of the tri-boson production process 
$p p \rightarrow e^{+}\nu_e \mu^{-} \bar{\nu}_{\mu} \gamma + X$ at $\sqrt{s} = 13.6\, {\rm TeV}$
reveals several key insights into the interplay between NLO QCD calculations and parton shower 
effects. The study of scale variations demonstrates that electroweak system observables,
such as the invariant mass, remain relatively stable under showering, with uncertainties
dominated by renormalization and factorization scale variations (approximately $\pm5\%$).
In contrast, jet observables show more pronounced shower effects and exhibit a dramatic 
reduction in uncertainties when moving from LO+PS to NLO+PS predictions, particularly in
the high-$p_T$ regime where variations decrease from $\approx70\%$ to $\approx 20\%$.

The investigation of migration effects has revealed significant implications for experimental
analyses and theoretical predictions. Our detailed study of generation-level cuts for photon
and lepton kinematics demonstrates that parton shower effects can lead to substantial migrations 
across cut boundaries, with impacts of up to $10\%$ on differential distributions - comparable
to or exceeding the scale variation uncertainties. These effects are particularly pronounced
near kinematic thresholds and become more significant with tighter rapidity cuts. The observed
migration patterns differ markedly between fixed-order and showered predictions, with the latter
showing systematic dependence on the generation-level cuts that must be carefully considered 
in precision studies.

Our findings demonstrate that NLO+PS matched calculations are essential
for precise predictions in multi-boson production processes, especially when 
studying observables sensitive to jet activity. Generation-level cuts must be 
chosen carefully, as they can substantially influence final distributions even 
after applying analysis-level cuts. The observation that migration effects can 
match or exceed scale variation uncertainties indicates these effects warrant 
systematic inclusion in uncertainty estimates.

The expanded capabilities of \Vbfnlo{} 3.0 through the BLHA interface open new avenues for
investigation in multi-boson physics. Future studies could extend this analysis to other
channels, investigate the impact on searches for anomalous gauge couplings, and compare 
results across different event generators and parton shower algorithms. Such comparisons 
would be particularly valuable for establishing the robustness of predictions in regions 
where migration effects are significant.

\section{Disclaimer}

The authors acknowledge the use of generative AI, specifically
Claude 3.5 Sonnet, in the preparation of this manuscript. This 
AI tool was employed to enhance the flow, clarity, and grammatical
quality of the text. While the scientific content, analysis, and
conclusions remain the sole work of the human authors, the AI assistance
has contributed to improving the overall readability and coherence
of the paper. The authors have carefully reviewed and verified all
AI-generated content to ensure its accuracy and alignment with the
intended scientific message.
\section{Acknowledgments}

F.C. and I.R acknowledge financial support by the Generalitat Valenciana, the spanish
goverment and ERDF funds from the European
Commission (Grants No. SEJI-2017/2017/019, CNS2022-136165, PID2023-151418NB-I00).

\bibliographystyle{unsrt}
\bibliography{triboson}

\end{document}